\documentclass[superscriptaddress,onecolumn,aps,pra,10pt]{revtex4-2}
\usepackage{amsmath,amsthm,amssymb,dsfont}
\usepackage{mdframed}
\usepackage{orcidlink}
\usepackage{CJK}
\usepackage{xcolor}
\newcommand{\ket}[1]{\left| #1 \right\rangle}
\newcommand{\bra}[1]{\left\langle #1 \right|}

\begin{document}

\title{The non-local Hong-Ou-Mandel effect}
\author{Yuki Kodama\,\orcidlink{0009-0004-1469-4549}}
\email{d265002@hiroshima-u.ac.jp}
\affiliation{Graduate School of Advanced Science and Engineering, Hiroshima University, Kagamiyama 1-3-1, Higashi Hiroshima 739-8530, Japan}
\author{Jonte R. Hance\,\orcidlink{0000-0001-8587-7618}}
\email{jonte.hance@newcastle.ac.uk}
\affiliation{School of Computing, Newcastle University, 1 Science Square, Newcastle upon Tyne, NE4 5TG, UK}
\affiliation{Quantum Engineering Technology Laboratories, Department of Electrical and Electronic Engineering, University of Bristol, Woodland Road, Bristol, BS8 1US, UK}
\author{Holger F. Hofmann\,\orcidlink{0000-0001-5649-9718}}
\email{hofmann@hiroshima-u.ac.jp}
\affiliation{Graduate School of Advanced Science and Engineering, Hiroshima University, Kagamiyama 1-3-1, Higashi Hiroshima 739-8530, Japan}

\begin{abstract}
Two-photon interference effects arise because photons are indistinguishable particles. In the well-known Hong-Ou-Mandel (HOM) effect, the transmission of two photons at a beam splitter interferes destructively with the reflection of both photons, requiring both photons to "bunch up" by leaving the beam splitter on the same side. Here, we show that the interference between locally propagating photons and photons exchanged by a mode swap can be implemented by post-selecting spatially separated photon outputs of a four-path interferometer. Even though the photons detected at spatially separated locations must have travelled along paths that never met up at the same beam splitter, the Hong-Ou-Mandel effect can be observed in correlations between the output ports that originate from the association of detection events with non-local output modes defined by the two single photon inputs. Local phase shifts can be used to map out non-classical correlations between the photons detected at different output locations, clarifying the role of linear optics in generating entanglement between spatially separated photons. Our work thus establishes a fundamental relation between multiphoton interference and entanglement, opening the door to new possibilities in optical quantum technologies.
\end{abstract}

\maketitle

\section{Introduction}
Multiphoton interference originates from the indistinguishability of photons in the bosonic Hilbert space of multiphoton wavefunctions \cite{mandel1999quantum,zou1991induced,santori2002indistinguishable,mandel1991coherence,lang2013correlations,menssen2017distinguishability,PhysRevA.91.013844,PhysRevLett.91.097902}. Having the advantage of being precisely controllable even under room-temperature conditions, multiphoton interference is a very versatile quantum resource \cite{pan2012multiphoton}. It has been applied to quantum computing \cite{RevModPhys.79.135,okamoto2009entanglement,flamini2018photonic,brod2019photonic,SA11,spring2013boson}, quantum communication\cite{NR07,Jeff08,RevModPhys.83.33,Aaronson:14} and quantum metrology\cite{giovannetti2011advances,polino2020photonic,dowling2008quantum}. Of particular significance might be the possibility of realizing universal quantum gates as shown by E. Knill, R. Laflamme, and G. J. Milburn \cite{KLM01}. Based on this observation, quantum computation can be realized using only linear optics, where the entangling operations are performed by multiphoton interference of inputs from single-photon sources and the results are obtained by precise photon counting in the output ports. Although the optical interference within the circuit becomes more complex as the number of photons increases, its underlying principle is the mode transformation that is defined by the interference of light field amplitudes at the linear optics elements that constitute the optical quantum circuit.

The phenomenon that best illustrates the principles of multiphoton interference is the Hong-Ou-Mandel (HOM) effect observed at a single beam splitter \cite{hong1987measurement}. When two photons are simultaneously incident on a beam splitter from opposite sides, they always emerge together on the same side, with one photon transmitted and the other reflected. Due to the indistinguishability of the photons, the reflection of both photons cannot be distinguished from the transmission of both photons. This indistinguishability results in destructive interferences between these two possibilities, causing the suppression of outputs on opposite sites of the beam splitter. It is interesting to note that the theoretical description of this effect is very similar to stimulated emission, where the probability of finding two photons in the same output mode is also enhanced. The intuitive image of such processes is that of a physical interaction by which photons appear to "stick" to each other, bunching up like two droplets that merge into a single larger object. However, multi-photon interference suggests no such interaction. Instead, the HOM effect arises from photon indistinguishability. This essential difference in the physics responsible for the effect can be tested by asking a simple question: is it possible to observe the HOM effect for two photons that never occupy the same path? If so, we can conclude that the HOM effect does not require any physical contact between the photons, indicating that multi-photon interference is intrinsically non-local.

In the conventional formulation of the HOM effect, the two photons are injected in two well-defined input paths that interfere at a single beam splitter. This limitation to only two optical modes associated with the paths that meet at the beam splitter suggests some form of physical interaction between the photons that travel along these paths. Consequently, two modes are not sufficient to demonstrate interference between photons that propagate along separate and distinct optical paths. To solve this problem, we introduce an interferometer with four modes, where photons are injected in only two modes, but can be detected in four different output modes. We can then separate the interferometer into two subsystems, $A$ and $B$, where entanglement between the two systems can be characterized by the photon statistics observed for different linear optics transformations applied to each local system \cite{wu2017evaluation,Kiyohara:20}. The HOM effect can then be realized by swapping the innermost pair of modes between $A$ and $B$. Interference occurs between the exchange of photons in the mode swap operation and the propagation of the photons along the paths that are not swapped. Beam splitters acting on the local input and the local output ensure that the two components are indistinguishable and post-selection removes the outcomes where both photons exit in the same local system. 

The state generated by this non-local HOM effect is an entangled state of two photons, where the output of interference between the two local modes of $A$ is correlated with the interference between the two local modes of $B$. It may seem counter intuitive that we identify this correlation of the outputs with the detection of both photons on the same side of a beam splitter in the conventional HOM effect. However, it can be shown that the two photons are detected in the same non-local output mode when the phase shifts of the interferences are properly aligned with each other. This non-local mode matching condition introduces an additional degree of freedom in the non-local HOM effect. By varying the phase shifts between the local modes in $A$ and in $B$, we can change the overlap between the non-local modes and the local output modes in which the photons are detected. Our results thus establish a fundamental relation between multiphoton interference and entanglement by introducing the concept of non-local modes as a means to characterize the intrinsic non-locality associated with optical coherence between remote systems. 

The structure of this paper is as follows. In Sec. \ref{sec:swap}, we introduce the mode-swapping operation in a four-mode interferometer and show how post-selection generates an entangled two-photon state between the two subsystems. In Sec. \ref{sec:modes}, we analyze the non-local HOM effect in terms of the mode matching between non-local output modes shared by the two subsystems and show how local phase shifts modify the observed output correlations. In Sec. \ref{sec:entanglement}, we show that the phase shifts correspond to local measurement settings and clarify the relation between the non-local HOM effect and the entanglement generated by mode swapping. In Sec. \ref{sec:conclusions}, we conclude the paper by highlighting the role of optical coherence between different paths in the generation and observation of non-classical photon number statistics.

\section{Entangling photons by mode swapping}
\label{sec:swap}

\begin{figure}[t]
  \centering
  \hspace*{-3cm}
  \includegraphics[width=0.6\linewidth]{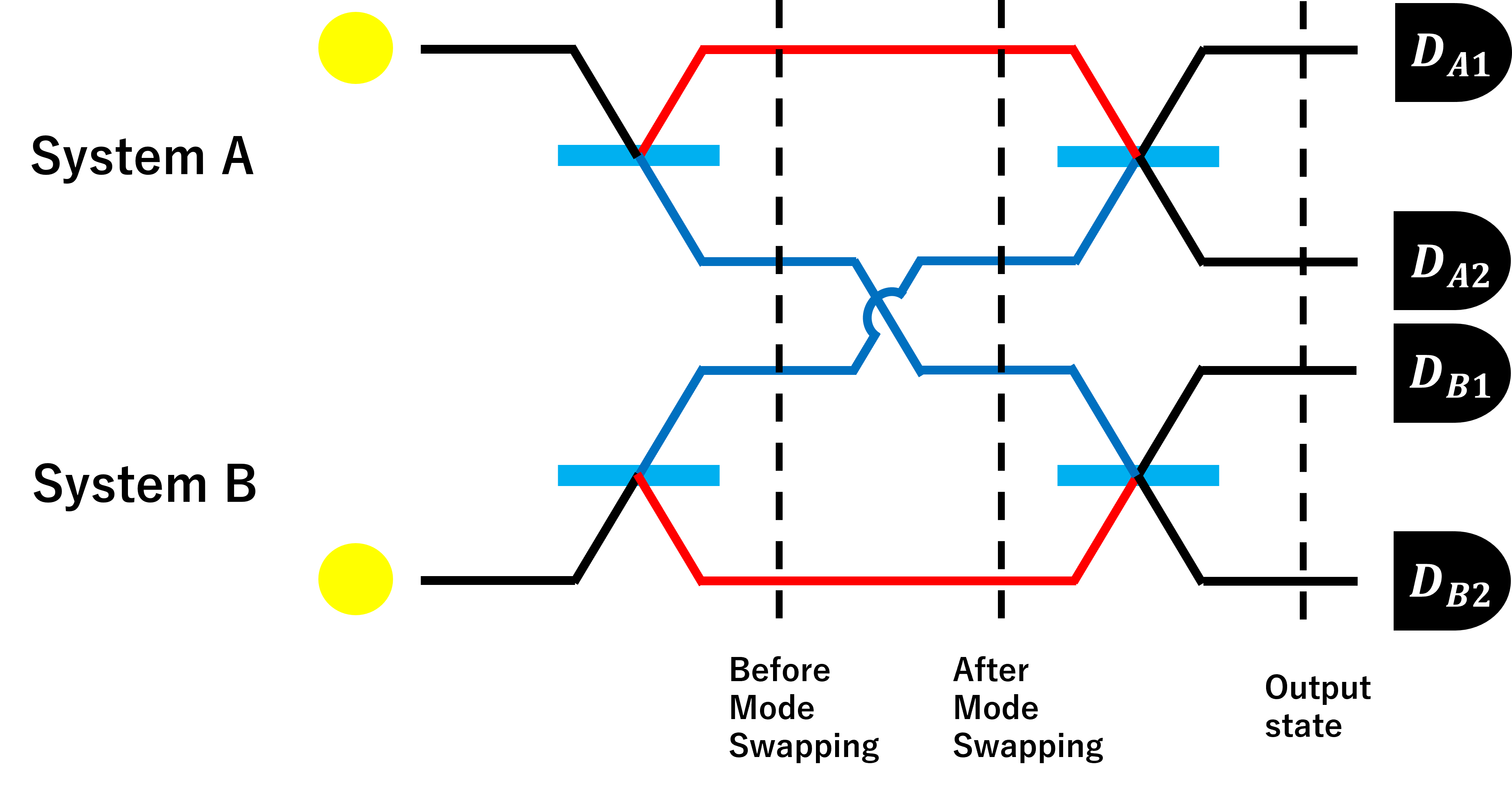}
  \caption{Setup for the implementation of the non-local Hong-Ou-Mandel effect. The setup is subdivided into two systems, $A$ and $B$, with two modes each. At the center, two of the modes are swapped, as described by a unitary transformation $\hat{U}_{\mathrm{swap}}$. To ensure equal superpositions of two photons in the swapped modes and two photons in the modes remaining on their side, the two input photons pass through beam splitters before the application of the mode swap. Likewise, beam splitters in the outputs ensure that the swapped modes and the remaining modes interfere in the final measurement.
  }
  \label{fig1}
\end{figure}

The Hong-Ou-Mandel effect can be traced back to the destructive interference between simultaneous reflection and simultaneous transmission of two photons at a single beam splitter. The reflection process and the transmission process both involve the same combination of input and output modes, so it is not possible to spatially separate them. In the following, we increase the number of modes to four, creating a spatial separation between two systems $A$ and $B$ with two modes each. For a single beam splitter, reflection means that the photons stay on the same side, while transmission means that they change sides. In the four mode system, we can implement an equivalent pair of processes by keeping one of the two modes in each system and swapping the other modes between the systems, as shown in Fig. \ref{fig1}. 

The well informed reader might be aware of previous experimental results on the implementation of non-local quantum interference effects by mode swapping \cite{qian2023multiphoton,wang2025violation}. It is important to note that the quantum interference effect investigated in these works involves nonlinear optics and requires the indistinguishability of photon pair creation processes. Although these results are without a doubt interesting in their own right, it is important to keep in mind that we are applying the mode swap operation to independently generated photons in well-defined input modes. This means that the quantum interference we intend to observe must arise entirely from the fundamental indistinguishability of photons as bosons. The two photons are initially distinguishable by the spatially separated optical modes that they were generated in. The mode swap itself is not sufficient to erase this distinguishability, since the swapped modes can still be distinguished from the ones that were not swapped. It is therefore necessary to apply single particle interference in both the input and the output, so that the origin of each photon cannot be distinguished in the output. This is the function of the beam splitters in the setup shown in Fig. \ref{fig1}. One photon each is injected into one input port of a beam splitter in both $A$ and $B$, resulting in a superposition of the two output ports for the states of each photon, as represented by the product state  
\begin{equation}
    \label{eq:input}
        \ket{\psi}=\frac{1}{\sqrt{2}}(i\ket{1,0}_A+ \ket{0,1}_A)\otimes\frac{1}{\sqrt{2}}(\ket{1,0}_B+i \ket{0,1}_B).
\end{equation}
Note that this is a two-photon four-mode state given in terms of the photon number basis of the modes. To represent the mode swap at the center of the setup shown in Fig. \ref{fig1}, we introduce the unitary operation $\hat{U}_{\mathrm{swap}}$ that performs a quantum swap operation on the two inner modes, resulting in entanglement between system $A$ and system $B$, 
\begin{equation}
    \label{eq:entangled}
           \hat{U}_{\mathrm{swap}} \ket{\psi}=\frac{1}{2}(i\ket{1,1;0,0}+\ket{0,1;1,0}-\ket{1,0;0,1}+i\ket{0,0;1,1}).
\end{equation}
Here, a semicolon is used to denote the separation between system $A$ and system $B$. 

Half of the time, both photons will end up in the same system. These components of the state in Eq.(\ref{eq:entangled}) do not contribute to non-local interference effects, so we will post-select only the outputs where one photon is found in each system. The normalized post-selected state is given by 
\begin{equation}
    \label{eq:postselect}
         \sqrt{2}\, \hat{\Pi} \, \hat{U}_{\mathrm{swap}} \ket{\psi}=\frac{1}{\sqrt{2}}(\ket{0,1;1,0}-\ket{1,0;0,1}),
\end{equation}
where $\hat{\Pi}$ is a projection onto the four dimensional subspace spanned by $\{\ket{1,0;1,0},\ket{1,0;0,1},\ket{0,1;1,0},\ket{0,1;0,1}\}$. The factor of $\sqrt{2}$ represents the normalization of the state given by $1/\sqrt{p}$, where $p=1/2$ is the post-selection probability. In the post-selected state, the photon in $A$ is maximally entangled with the photon in $B$. This means that any detection of the photon in $A$ will be correlated to a corresponding detection event in $B$. The mode swap operation has entangled the two photons even though they never physically occupied the same space. 

To confirm the two photon interference between swapped photons and photons remaining in the same system, we use the output beam splitters to perform a measurement of equal superpositions of the swapped modes and the remaining modes. These measurements can be represented by state vectors $\bra{D_i}$, where the bra notation is used to indicate that the state represents a specific measurement outcome. The single photon outcomes in system $A$ are given by
\begin{eqnarray}
\bra{D_{A1}} &=& \frac{1}{\sqrt{2}} \left(i \bra{1,0}_A + \bra{0,1}_A\right)  
\nonumber \\ 
\bra{D_{A2}} &=& \frac{1}{\sqrt{2}} \left(\bra{1,0}_A + i \bra{0,1}_A\right),  
\end{eqnarray}
and the single photon outcomes in system $B$ are given by 
\begin{eqnarray}
\bra{D_{B1}} &=& \frac{1}{\sqrt{2}} \left(i \bra{1,0}_B + \bra{0,1}_B\right),  
\nonumber \\ 
\bra{D_{B2}} &=& \frac{1}{\sqrt{2}} \left(\bra{1,0}_B + i \bra{0,1}_B\right).  
\end{eqnarray}
The detection basis can be used to express the output state as
\begin{equation}
\label{eq:detect}
         \sqrt{2}\, \hat{\Pi}\, \hat{U}_{\mathrm{swap}} \ket{\psi}=\frac{-1}{\sqrt{2}}(\ket{D_{A2};D_{B1}}-\ket{D_{A1};D_{B2}}).
\end{equation}
As expected for a maximally entangled state, the detection events are correlated, with $P(D_{A1},D_{B1})$=0 and $P(D_{A2},D_{B2})=0$. Similar to the conventional HOM effect, destructive interference between the swapped photon pair and the pair of photons that stayed in their respective systems reduces these two probabilities to zero. The two remaining outcomes should then correspond to photons exiting the beam splitter on the same side. In the present scenario, these two outcomes are given by the non-zero outcome probabilities
\begin{eqnarray}
P(D_{A2},D_{B1})&=&1/2,
\nonumber \\
P(D_{A1},D_{B2})&=&1/2.
\end{eqnarray}
The non-local HOM effect thus associates $D_{A2}$ and $D_{B1}$ with one ``side'' of the output, and $D_{A1}$ and $D_{B2}$ with the other. As we will show in the following, this association can be explained in terms of the non-local output modes corresponding to the input modes of the two photons.

\section{Mode Matching in the Non-Local Hong–Ou–Mandel Effect}\label{sec:modes}

In this section, we show that the two-photon interference implemented by the four-path interferometer in Fig. \ref{fig1} is a non-local realization of the two-mode HOM effect. The addition of two empty modes adds a degree of freedom to the output measurements that can be explored by applying local phase shifts to the output modes. It is shown that the phase shifts select the interfering modes, modifying the HOM effect by generating a well controlled mode mismatch between the output modes on both sides. 

Since the two photons are injected separately into the input modes $\hat{a}_{A1}$ and $\hat{a}_{B2}$, the initial state can be expressed using the creation operators of these modes,
\begin{equation}
    \label{eq:InitialState}
    \ket{\psi}=\hat{a}_{A1}^\dagger \hat{a}_{B2}^\dagger \ket{\mathrm{vac.}}.
\end{equation}
The non-local HOM effect is realized by separating the input modes at the beam splitter, followed by a mode swap of the innermost modes. Each of the two input modes $\hat{a}_{A1}$ and $\hat{a}_{B2}$ is transmitted to each of the output modes $\hat{b}_i$ with a transmittance of $1/4$. The creation operators can be expressed as
\begin{align}
\label{eq:modetransmission}
    \hat{a}_{A1}^{\dagger}(0) &= \frac{i}{2}\bigl(\hat{b}_{A2}^{\dagger}+\hat{b}_{B1}^{\dagger}\bigr)
    - \frac{1}{2}\bigl(\hat{b}_{A1}^{\dagger}-\hat{b}_{B2}^{\dagger}\bigr),\nonumber \\[6pt]
    \hat{a}_{B2}^{\dagger}(0) &= \frac{i}{2}\bigl(\hat{b}_{A2}^{\dagger}+\hat{b}_{B1}^{\dagger}\bigr)
    + \frac{1}{2}\bigl(\hat{b}_{A1}^{\dagger}-\hat{b}_{B2}^{\dagger}\bigr).
\end{align}
As indicated by the brackets, there are only two output modes, $(\hat{b}_{A2}+\hat{b}_{B1})/\sqrt{2}$ and $(\hat{b}_{A1}-\hat{b}_{B2})/\sqrt{2}$. However, each of these output modes is delocalized, with one component in $A$ and the other in $B$. The output photon statistics can be written as
\begin{equation}
\label{eq:nonlocalHOM}
\begin{aligned}
    \hat{a}_{A1}^{\dagger}(0)\hat{a}_{B2}^{\dagger}(0) \ket{\mathrm{vac.}} &= -\frac{1}{4}\bigl(\hat{b}_{A2}^{\dagger}+\hat{b}_{B1}^{\dagger}\bigr)^2 \ket{\mathrm{vac.}}
    - \frac{1}{4}\bigl(\hat{b}_{A1}^{\dagger}-\hat{b}_{B2}^{\dagger}\bigr)^2 \ket{\mathrm{vac.}},
\end{aligned}
\end{equation}
where both photons exit in the same non-local mode \cite{SPIE}. Even if the photons are detected in different systems, they are still found in the same mode, where $\ket{D_{A2}}$ detects the $\hat{b}_{A2}$ component of mode $(\hat{b}_{A2}+\hat{b}_{B1})/\sqrt{2}$ in $A$ and $\ket{D_{B1}}$ detects the $\hat{b}_{B1}$ component of the same mode in $B$. The non-locality of modes thus explains the correlations between local photon detections. 

\begin{figure}[t] % t=上, b=下, h=近く, !で強め
  \centering
  \includegraphics[width=\linewidth]{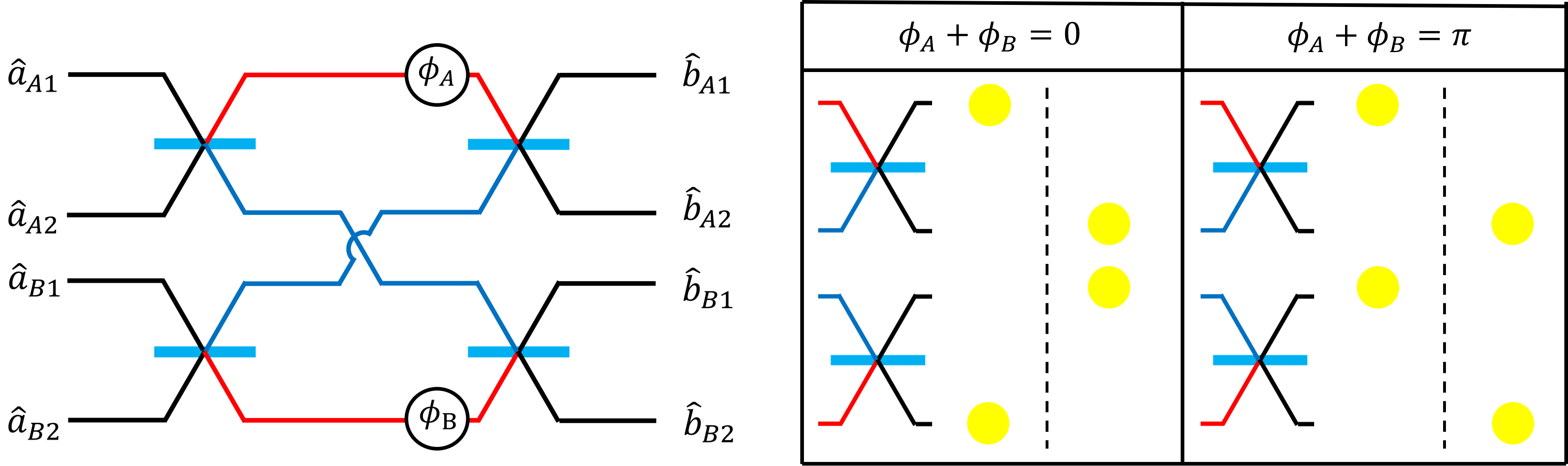} % 拡張子は省略可
  \caption{Effects of local phase shifts on the non-local HOM effect. Input mode $\hat{a}_{A1}$ and input mode$\hat{a}_{B2}$ receive one photon each. Local phase shifts $\phi_A$ and $\phi_B$ are applied to the outer modes, ensuring that the phase shifts are localized in both the input and the output. For a phase shift of $\phi_A+\phi_B=0$, the non-local HOM effect is not affected by the phases. For $\phi_A+\phi_B=\pi$, the non-local HOM effect occurs between different non-local modes, resulting in a different correlation between the output modes.}
  \label{fig2}
\end{figure}

%%--figure needs to be changed. Remove (a) (b), label only the input modes used, a_A1 and a_B2, change output mode labels to b_A1, b_A2, b_B1, b_B2

We can now modify the non-local modes by introducing phase shifts into the outermost paths as shown in Fig. \ref{fig2}. These phase shifts change the transmission of input modes, with the phase shift $\phi_A$ acting on input mode $\hat{a}_{A1}$ and the phase shift $\phi_B$ acting on input mode $\hat{a}_{B2}$,
\begin{align}
\label{eq:phaseshift}
    \hat{a}_{A1}^{\dagger}(\phi_A) &= \frac{i}{2}\bigl(e^{i\phi_{A}}\hat{b}_{A2}^{\dagger}+\hat{b}_{B1}^{\dagger}\bigr)
    - \frac{1}{2}e^{i\phi_{A}}\bigl(\hat{b}_{A1}^{\dagger}-e^{-i\phi_{A}}\hat{b}_{B2}^{\dagger}\bigr) ,\nonumber \\[6pt]
    \hat{a}_{B2}^{\dagger} (\phi_B)&= \frac{i}{2}e^{i\phi_{B}}\bigl(e^{-i\phi_{B}}\hat{b}_{A2}^{\dagger}+\hat{b}_{B1}^{\dagger}\bigr)
    + \frac{1}{2}\bigl(\hat{b}_{A1}^{\dagger}-e^{i\phi_{B}}\hat{b}_{B2}^{\dagger}\bigr)
\end{align}
The non-local modes in $\hat{a}_{A1}$ only match the non-local modes in $\hat{a}_{B2}$ when the phase shifts are equal but opposite, $\phi_B=-\phi_A$. In this case, the non-local HOM effect is observed as above, with an unobservable phase shift added to the coherence between the mode components in $A$ and in $B$,
\begin{equation}
\label{eq:nlHOM2}
\begin{aligned}
    \hat{a}_{A1}^{\dagger}(\phi)\hat{a}_{B2}^{\dagger}(-\phi) \ket{\mathrm{vac.}} &= -\frac{1}{4}\bigl(e^{i\phi}\hat{b}_{A2}^{\dagger}+\hat{b}_{B1}^{\dagger}\bigr)^2 \ket{\mathrm{vac.}}
    - \frac{1}{4}\bigl(\hat{b}_{A1}^{\dagger}-e^{-i\phi}\hat{b}_{B2}^{\dagger}\bigr)^2 \ket{\mathrm{vac.}}.
\end{aligned}
\end{equation}
When the sum of the phase shifts is not zero, it is necessary to consider the mode mismatch between the non-local output modes associated with the input modes $\hat{a}_{A1}$ and $\hat{a}_{B2}$. The non-local modes can be identified in Eq. (\ref{eq:phaseshift}). Mode matching can be defined in terms of the single photon overlap between the modes. According to Eqs.(\ref{eq:nonlocalHOM}) and (\ref{eq:nlHOM2}), we can distinguish the non-local modes formed by the inner output modes $\hat{b}_{A2}$ and $\hat{b}_{B1}$ from the non-local modes formed by the outer output modes $\hat{b}_{A1}$ and $\hat{b}_{B2}$. The mode matching of the inner and outer output modes of $\hat{a}_{A1}$ and $\hat{a}_{B2}$ can then be given by the overlap of their hypothetical single photon states,
\begin{align}
    \label{eq:modematch}
        |\bra{\mathrm{vac.}}\frac{1}{\sqrt{2}}\bigl(e^{-i\phi_{A}}\hat{b}_{A2}+\hat{b}_{B1}\bigl)\frac{1}{\sqrt{2}}\bigl(e^{-i\phi_{B}}\hat{b}_{A2}^{\dagger}+\hat{b}_{B1}^{\dagger}\bigl)\ket{\mathrm{vac.}}|^{2}&=\frac{1}{2}\bigl(1+\cos\bigl(\phi_{A}+\phi_{B}\bigl)\bigl)
\nonumber \\
        |\bra{\mathrm{vac.}}\frac{1}{\sqrt{2}}\bigl(\hat{b}_{A1}-e^{i\phi_{A}}\hat{b}_{B2}\bigl)\frac{1}{\sqrt{2}}\bigl(\hat{b}_{A1}^\dagger-e^{i\phi_{B}}\hat{b}_{B2}^\dagger\bigl)\ket{\mathrm{vac.}}|^{2}&=\frac{1}{2}\bigl(1+\cos\bigl(\phi_{A}+\phi_{B}\bigl)\bigl)
\end{align}
We can now apply mode matching to determine the output probabilities of the non-local HOM effect. Using the mode relations shown in Eq.(\ref{eq:phaseshift}), the post-selected probabilities are defined by 
\begin{align}
    P(D_{Ai},D_{Bj}) &= 2 |\bra{\mathrm{vac.}} \hat{b}_{Ai}\hat{b}_{Bj} \hat{a}^\dagger_{A1} \hat{a}^\dagger_{B2} \ket{\mathrm{vac.}}|^2.  
\end{align} 
Note that the post-selection probability is always $1/2$, since that is the probability of finding both photons on opposite sides after the mode swap.
The specific probabilities are determined from the interferences between the amplitude products corresponding to $\hat{b}_{Ai}$ from $\hat{a}_{A1}$ and $\hat{b}_{Bj}$ from $\hat{a}_{A2}$ (no photons in the swapped modes), and to $\hat{b}_{Ai}$ from $\hat{a}_{B2}$ and $\hat{b}_{Bj}$ from $\hat{a}_{A1}$ (both photons in the swapped mode). The interference is the same as the one that appears in the single photon mode matching relation given in Eq.(\ref{eq:modematch}). The probabilities of the four output combinations obtained when one photon is in $A$ and one photon is in $B$ are
\begin{align}
\label{eq:outputphase}
    P(D_{A1},D_{B1}) = \frac{1}{4}\bigl(1-\cos\bigl(\phi_{A}+\phi_{B}\bigl)\bigl) \nonumber \\[6pt]
    P(D_{A1},D_{B2}) = \frac{1}{4}\bigl(1+\cos\bigl(\phi_{A}+\phi_{B}\bigl)\bigl) \nonumber \\[6pt]
    P(D_{A2},D_{B1}) = \frac{1}{4}\bigl(1+\cos\bigl(\phi_{A}+\phi_{B}\bigl)\bigl) \nonumber \\[6pt]
    P(D_{A2},D_{B2}) = \frac{1}{4}\bigl(1-\cos\bigl(\phi_{A}+\phi_{B}\bigl)\bigl).
\end{align}
At $\phi_A+\phi_B=\pi$, the probabilities of $(D_{A1},D_{B1})$ and $(D_{A2},D_{B2})$ are $1/2$ and the non-local modes are given by 
\begin{align}
    \hat{a}^\dagger_{A1}(\phi) &\ = -\frac{1}{2}\left(e^{i \phi}\hat{b}^\dagger_{A1}-i \, \hat{b}^\dagger_{B1} \right) + \frac{i}{2}e^{i \phi} \left(\hat{b}^\dagger_{A2}-i \, e^{-i\phi} \hat{b}^\dagger_{B2}\right)
    \nonumber \\
    \hat{a}^\dagger_{B2}(\pi-\phi) &\ = \frac{1}{2} e^{-i\phi}\left(e^{i \phi} \hat{b}^\dagger_{A1} -i \hat{b}^\dagger_{B1}\right)+\frac{i}{2}\left(\hat{b}^\dagger_{A2}-i \, e^{-i \phi}\hat{b}^\dagger_{B2}\right)
\end{align}
Local phase shifts can switch the output correlations of the two photons from correlations between $D_{A1}$ and $D_{B2}$ ($D_{A2}$ and $D_{B1}$) to $D_{A1}$ and $D_{B1}$ ($D_{A2}$ and $D_{B2}$). If the phase shifts are considered as a part of a local measurement process, it is possible to identify the mode matching effects of the non-local HOM effect with the non-local correlations between the photon in $A$ and the photon in $B$. The generation of entanglement by mode swapping can then be explained in terms of the non-local HOM effect of the input modes $\hat{a}_{A1}$ and $\hat{a}_{B2}$.

\section{Relation between the Hong-Ou-Mandel effect and entanglement}\label{sec:entanglement}

As we have seen in Sec. \ref{sec:swap} above, the post-selected state after the mode swap is a maximally entangled state. Before the interference at the output beam splitters, the modes can be separated into the swapped modes $\hat{b}_{S\!A}$ and $\hat{b}_{S\!B}$ and the unswapped outer modes $\hat{b}_{U\!A}$ and $\hat{b}_{U\!B}$. The input modes are then given by 
\begin{align}
\label{eq:mid}
    \hat{a}^\dagger_{A1} &= \frac{i}{\sqrt{2}}\left(\hat{b}^\dagger_{U\!A} - i \,\hat{b}^\dagger_{S\!B}\right) \nonumber \\
    \hat{a}^\dagger_{B2} &= \frac{1}{\sqrt{2}}\left(\hat{b}^\dagger_{S\!A} + i \,\hat{b}^\dagger_{U\!B}\right).
 \end{align}
The phase shifts can now be included in the description of the modes identified by each detector,
\begin{align}
\label{eq:Dmodes}
    \bra{D_{A1}(\phi_{A})} &= \frac{i}{\sqrt{2}}\left(\,e^{i\phi_{A}}\, \bra{\mathrm{vac.}}_A\,\hat{b}_{U\!A}- i\, \bra{\mathrm{vac.}}_{A}\,\hat{b}_{S\!A} \right) ,\nonumber \\[6pt]
    \bra{D_{A2}(\phi_{A})} &= \frac{1}{\sqrt{2}}\left(\,e^{i\phi_{A}}\, \bra{\mathrm{vac.}}_{A}\,\hat{b}_{U\!A} +i \, \bra{\mathrm{vac.}}_{A}\,\hat{b}_{S\!A}\right), \nonumber \\[6pt]
    \bra{D_{B1}(\phi_{B})} &= \frac{1}{\sqrt{2}}\left(\,e^{i\phi_{B}} \,\bra{\mathrm{vac.}}_{B}\,\hat{b}_{U\!B}+ i \bra{\mathrm{vac.}}_B\,\hat{b}_{S\!B} \right) ,\nonumber \\[6pt]
    \bra{D_{B2}(\phi_{B})} &= \frac{i}{\sqrt{2}}\left(\,e^{i\phi_{B}}\,\bra{\mathrm{vac.}}_B\,\hat{b}_{\mathrm{UB}}-i\,\bra{\mathrm{vac.}}_B\,\hat{b}_{\mathrm{SB}}\right).
\end{align}
Each detector mode is represented by local interferences between the two input modes, $\hat{a}_{A1}$ and $\hat{a}_{B2}$. The non-local HOM effect is observed if the interferences in $A$ correspond to those in $B$. For every combination of $D_{Ai}$ and $D_{Bj}$, the contributions with creation operators $\hat{b}^\dagger_{U\!A}$ and $\hat{b}^\dagger_{U\!B}$ interfere with the contribution with $\hat{b}_{S\!A}$ and $\hat{b}_{S\!B}$. For $D_{A1}$ and $D_{B2}$ and for $D_{A2}$ and $D_{B1}$, the phase difference is $\phi_A+\phi_B$. For $D_{A1}$ and $D_{B1}$ and for $D_{A2}$ and $D_{B2}$, it is $\phi_A+\phi_B+\pi$.

Eq.(\ref{eq:postselect}) shows that the post selected output state is a maximally entangled state of the single photon Hilbert spaces in $A$ and in $B$. Can the non-locality of modes be used to explain the origins of this entanglement? As Eq.(\ref{eq:mid}) shows, the mode swap delocalizes the input modes by splitting them into components that are subsequently distributed to different locations. From the viewpoint of wave optics, the non-local HOM effect is observed because the phase relations between the two input modes determine the interferences in both $A$ and $B$ in a similar manner. This is the reason why maximal correlations are observed at $\phi_A+\phi_B=0$ and at $\phi_A+\phi_B=\pi$. Eq.(\ref{eq:Dmodes}) shows that the different phases $\phi_A$ and $\phi_B$ represent different measurement settings. This means that the phase dependence of the non-local HOM effect can be used to illustrate quantum steering \cite{PhysRevLett.98.140402}, where a measurement of $D_{A1}$ prepares a single photon state in $B$ defined by the local phase $\phi_A$ in system $A$.
\begin{align}
   _A \bra{D_{A1}(\phi_A)} \hat{a}^\dagger_{A1} \hat{a}^\dagger_{B2} \ket{\mathrm{vac.}}_{AB} &= \frac{-i}{2\sqrt{2}} \left(e^{i\phi_A} \hat{b}_{U\!B}^\dagger + i \hat{b}^\dagger_{S\!B} \right) \ket{\mathrm{vac.}}
\end{align}
The detection of a photon in $D_{A1}$ in $A$ determines the phase relation between $\hat{b}_{U\!A}$ and $\hat{b}_{S\!A}$ in $A$. Since there are only two input modes, this phase relation originates from the phase relation between the modes $\hat{a}_{A1}$ and $\hat{a}_{B2}$. The same phase relation necessarily determines the interference between $\hat{b}_{S\!B}$ and $\hat{b}_{U\!B}$ in $B$. The non-local HOM effect thus explains the generation of entanglement in multi-photon multi-mode systems as a result of the phase coherence of non-local modes with well defined input photon numbers. Quantum correlations are explained by the observation of identical phase differences between pairs of non-local input modes, where photons detected in $A$ must always show the same interference patterns as photons in $B$. We have thus shown that the natural non-locality of optical modes results in a non-local version of the HOM effect, where the correlation between the output ports at which the spatially separated photons are detected confirms the generation of quantum entanglement.

\section{Conclusions}
\label{sec:conclusions}

We have demonstrated the intrinsic non-locality of the HOM effect by embedding the underlying multi-photon interference effect in a four mode interferometer, spatially separating the local transmission of photons from the exchange of photons between the two sub-systems of the interferometer. Our analysis highlights the fundamental difference between multiphoton interference and the interactions between photons in nonlinear optical media. While interactions between photons are inherently local, involving the local nonlinear response of a medium, multiphoton interference is inherently non-local due to the natural non-locality of optical modes that are defined by the optical coherence between different paths. When photon number states are represented by creation and annihilation operators, this relation between coherence and non-locality is expressed in its most natural form. It may well be that the intuitive description of photons as particles is distracting us from the more fundamental physics expressed by the optical interferences between these operators, where the mathematical creation of a photon in a mode actually references the field coherence of this mode. The analysis of the non-local HOM effect shows how the seemingly classical coherence of fields explains the quantum non-locality observed in the photon statistics at the output ports of a four path interferometer. This is by no means trivial, since it is generally assumed that classical coherence only corresponds to single photon statistics. Our analysis may thus serve as a starting point for a more systematic exploration of non-classical effects in quantum optical circuits involving large numbers of photons.

\section*{Acknowledgements} 
This work was supported by JST ERATO Grant Number JPMJER2402. JRH acknowledges support from a Royal Society Research Grant (RG/R1/251590) and from their EPSRC Quantum Technologies Career Acceleration Fellowship (UKRI1217).

\nocite{*}
\bibliography{ref.bib}

@article{giovannetti2011advances,
  title={Advances in quantum metrology},
  author={Giovannetti, Vittorio and Lloyd, Seth and Maccone, Lorenzo},
  journal={Nature photonics},
  volume={5},
  number={4},
  pages={222--229},
  year={2011},
  publisher={Nature Publishing Group UK London}
}

@article{spring2013boson,
  title={Boson sampling on a photonic chip},
  author={Spring, Justin B and Metcalf, Benjamin J and Humphreys, Peter C and Kolthammer, W Steven and Jin, Xian-Min and Barbieri, Marco and Datta, Animesh and Thomas-Peter, Nicholas and Langford, Nathan K and Kundys, Dmytro and others},
  journal={Science},
  volume={339},
  number={6121},
  pages={798--801},
  year={2013},
  publisher={American Association for the Advancement of Science}
}

@article{mandel1991coherence,
  title={Coherence and indistinguishability},
  author={Mandel, Leonard},
  journal={Optics letters},
  volume={16},
  number={23},
  pages={1882--1883},
  year={1991},
  publisher={Optical Society of America}
}

@article{lang2013correlations,
  title={Correlations, indistinguishability and entanglement in Hong--Ou--Mandel experiments at microwave frequencies},
  author={Lang, C and Eichler, Christopher and Steffen, L and Fink, JM and Woolley, Matthew J and Blais, Alexandre and Wallraff, Andreas},
  journal={Nature Physics},
  volume={9},
  number={6},
  pages={345--348},
  year={2013},
  publisher={Nature Publishing Group UK London}
}

@article{menssen2017distinguishability,
  title={Distinguishability and many-particle interference},
  author={Menssen, Adrian J and Jones, Alex E and Metcalf, Benjamin J and Tichy, Malte C and Barz, Stefanie and Kolthammer, W Steven and Walmsley, Ian A},
  journal={Physical review letters},
  volume={118},
  number={15},
  pages={153603},
  year={2017},
  publisher={APS}
}

@article{okamoto2009entanglement,
  title={An entanglement filter},
  author={Okamoto, Ryo and O'Brien, Jeremy L and Hofmann, Holger F and Nagata, Tomohisa and Sasaki, Keiji and Takeuchi, Shigeki},
  journal={Science},
  volume={323},
  number={5913},
  pages={483--485},
  year={2009},
  publisher={American Association for the Advancement of Science}
}

@article{PhysRevLett.91.097902,
  title = {Entanglement of Indistinguishable Particles Shared between Two Parties},
  author = {Wiseman, H. M. and Vaccaro, John A.},
  journal = {Phys. Rev. Lett.},
  volume = {91},
  issue = {9},
  pages = {097902},
  numpages = {4},
  year = {2003},
  month = {Aug},
  publisher = {American Physical Society},
  doi = {10.1103/PhysRevLett.91.097902},
  url = {https://link.aps.org/doi/10.1103/PhysRevLett.91.097902}
}

@article{santori2002indistinguishable,
  title={Indistinguishable photons from a single-photon device},
  author={Santori, Charles and Fattal, David and Vu{\v{c}}kovi{\'c}, Jelena and Solomon, Glenn S and Yamamoto, Yoshihisa},
  journal={nature},
  volume={419},
  number={6907},
  pages={594--597},
  year={2002},
  publisher={Nature Publishing Group UK London}
}

@article{zou1991induced,
  title={Induced coherence and indistinguishability in optical interference},
  author={Zou, Xing-Yu and Wang, Lei J and Mandel, Leonard},
  journal={Physical review letters},
  volume={67},
  number={3},
  pages={318},
  year={1991},
  publisher={APS}
}

@article{RevModPhys.79.135,
  title = {Linear optical quantum computing with photonic qubits},
  author = {Kok, Pieter and Munro, W. J. and Nemoto, Kae and Ralph, T. C. and Dowling, Jonathan P. and Milburn, G. J.},
  journal = {Rev. Mod. Phys.},
  volume = {79},
  issue = {1},
  pages = {135--174},
  numpages = {0},
  year = {2007},
  month = {Jan},
  publisher = {American Physical Society},
  doi = {10.1103/RevModPhys.79.135},
  url = {https://link.aps.org/doi/10.1103/RevModPhys.79.135}
}

@article{hong1987measurement,
  title={Measurement of subpicosecond time intervals between two photons by interference},
  author={Hong, Chong-Ki and Ou, Zhe-Yu and Mandel, Leonard},
  journal={Physical review letters},
  volume={59},
  number={18},
  pages={2044},
  year={1987},
  publisher={APS}
}

@article{pan2012multiphoton,
  title={Multiphoton entanglement and interferometry},
  author={Pan, Jian-Wei and Chen, Zeng-Bing and Lu, Chao-Yang and Weinfurter, Harald and Zeilinger, Anton and {\.Z}ukowski, Marek},
  journal={Reviews of Modern Physics},
  volume={84},
  number={2},
  pages={777--838},
  year={2012},
  publisher={APS}
}

@article{qian2023multiphoton,
  title={Multiphoton non-local quantum interference controlled by an undetected photon},
  author={Qian, Kaiyi and Wang, Kai and Chen, Leizhen and Hou, Zhaohua and Krenn, Mario and Zhu, Shining and Ma, Xiao-song},
  journal={Nature Communications},
  volume={14},
  number={1},
  pages={1480},
  year={2023},
  publisher={Nature Publishing Group UK London}
}

@article{PhysRevA.91.013844,
  title = {Partial indistinguishability theory for multiphoton experiments in multiport devices},
  author = {Shchesnovich, V. S.},
  journal = {Phys. Rev. A},
  volume = {91},
  issue = {1},
  pages = {013844},
  numpages = {16},
  year = {2015},
  month = {Jan},
  publisher = {American Physical Society},
  doi = {10.1103/PhysRevA.91.013844},
  url = {https://link.aps.org/doi/10.1103/PhysRevA.91.013844}
}

@article{brod2019photonic,
  title={Photonic implementation of boson sampling: a review},
  author={Brod, Daniel J and Galv{\~a}o, Ernesto F and Crespi, Andrea and Osellame, Roberto and Spagnolo, Nicol{\`o} and Sciarrino, Fabio},
  journal={Advanced Photonics},
  volume={1},
  number={3},
  pages={034001--034001},
  year={2019},
  publisher={Society of Photo-Optical Instrumentation Engineers}
}

@article{flamini2018photonic,
  title={Photonic quantum information processing: a review},
  author={Flamini, Fulvio and Spagnolo, Nicolo and Sciarrino, Fabio},
  journal={Reports on Progress in Physics},
  volume={82},
  number={1},
  pages={016001},
  year={2018},
  publisher={IOP Publishing}
}

@article{polino2020photonic,
  title={Photonic quantum metrology},
  author={Polino, Emanuele and Valeri, Mauro and Spagnolo, Nicol{\`o} and Sciarrino, Fabio},
  journal={AVS Quantum Science},
  volume={2},
  number={2},
  year={2020},
  publisher={AIP Publishing}
}

@article{NR07,
  title={Quantum communication},
  author={Gisin, Nicolas and Thew, Rob},
  journal={Nature photonics},
  volume={1},
  number={3},
  pages={165--171},
  year={2007},
  publisher={Nature Publishing Group UK London}
}

@article{mandel1999quantum,
  title={Quantum effects in one-photon and two-photon interference},
  author={Mandel, Leonard},
  journal={Reviews of Modern Physics},
  volume={71},
  number={2},
  pages={S274},
  year={1999},
  publisher={APS}
}

@article{Jeff08,
  title={The quantum internet},
  author={Kimble, H Jeff},
  journal={Nature},
  volume={453},
  number={7198},
  pages={1023--1030},
  year={2008},
  publisher={Nature Publishing Group}
}

@article{RevModPhys.83.33,
  title = {Quantum repeaters based on atomic ensembles and linear optics},
  author = {Sangouard, Nicolas and Simon, Christoph and de Riedmatten, Hugues and Gisin, Nicolas},
  journal = {Rev. Mod. Phys.},
  volume = {83},
  issue = {1},
  pages = {33--80},
  numpages = {0},
  year = {2011},
  month = {Mar},
  publisher = {American Physical Society},
  doi = {10.1103/RevModPhys.83.33},
  url = {https://link.aps.org/doi/10.1103/RevModPhys.83.33}
}

@inproceedings{Aaronson:14,
author = {Scott Aaronson and Alex Arkhipov},
booktitle = {Research in Optical Sciences},
journal = {Research in Optical Sciences},
pages = {QTh1A.2},
publisher = {Optica Publishing Group},
title = {The Computational Complexity of Linear Optics},
year = {2014},
url = {https://opg.optica.org/abstract.cfm?URI=QIM-2014-QTh1A.2},
doi = {10.1364/QIM.2014.QTh1A.2},
}

@article{KLM01,
  title={A scheme for efficient quantum computation with linear optics},
  author={Knill, Emanuel and Laflamme, Raymond and Milburn, Gerald J},
  journal={nature},
  volume={409},
  number={6816},
  pages={46--52},
  year={2001},
  publisher={Nature Publishing Group UK London}
}

@article{wang2025violation,
  title={Violation of Bell inequality with unentangled photons},
  author={Wang, Kai and Hou, Zhaohua and Qian, Kaiyi and Chen, Leizhen and Krenn, Mario and Aspelmeyer, Markus and Zeilinger, Anton and Zhu, Shining and Ma, Xiao-Song},
  journal={Science Advances},
  volume={11},
  number={31},
  pages={eadr1794},
  year={2025},
  publisher={American Association for the Advancement of Science}
}

@inproceedings{SA11,
  title={The computational complexity of linear optics},
  author={Aaronson, Scott and Arkhipov, Alex},
  booktitle={Proceedings of the forty-third annual ACM symposium on Theory of computing},
  pages={333--342},
  year={2011}
}

@article{wu2017evaluation,
  title={Evaluation of bipartite entanglement between two optical multi-mode systems using mode translation symmetry},
  author={Wu, Jun-Yi and Hofmann, Holger F},
  journal={New Journal of Physics},
  volume={19},
  number={10},
  pages={103032},
  year={2017},
  publisher={IOP Publishing}
}

@article{Kiyohara:20,
author = {Takayuki Kiyohara and Naoki Yamashiro and Ryo Okamoto and Hirotaka Araki and Jun-Yi Wu and Holger F. Hofmann and Shigeki Takeuchi},
journal = {Optica},
number = {11},
pages = {1517--1523},
publisher = {Optica Publishing Group},
title = {Direct and efficient verification of entanglement between two multimode--multiphoton systems},
volume = {7},
month = {Nov},
year = {2020},
url = {https://opg.optica.org/optica/abstract.cfm?URI=optica-7-11-1517},
doi = {10.1364/OPTICA.397943},
}

@article{PhysRevLett.98.140402,
  title = {Steering, Entanglement, Nonlocality, and the Einstein-Podolsky-Rosen Paradox},
  author = {Wiseman, H. M. and Jones, S. J. and Doherty, A. C.},
  journal = {Phys. Rev. Lett.},
  volume = {98},
  issue = {14},
  pages = {140402},
  numpages = {4},
  year = {2007},
  month = {Apr},
  publisher = {American Physical Society},
  doi = {10.1103/PhysRevLett.98.140402},
  url = {https://link.aps.org/doi/10.1103/PhysRevLett.98.140402}
}

@article{dowling2008quantum,
  title={Quantum optical metrology--the lowdown on high-N00N states},
  author={Dowling, Jonathan P},
  journal={Contemporary physics},
  volume={49},
  number={2},
  pages={125--143},
  year={2008},
  publisher={Taylor \& Francis}
}

@article{SPIE,
  title={Implementation of nonlocal multi-photon interference
by mode swapping},
  author={Holger F. Hofmann and Yuki Kodama and Jonte R. Hance},
  journal={Proc. SPIE},
  volume={13618},
  pages={136180H},
  year={2025},
  doi = {10.1117/12.3063247}
}
\end{document}